\documentclass{svproc}
\usepackage{url}

\usepackage[breaklinks=true]{hyperref}

\usepackage{graphicx}
\usepackage{xcolor}
\usepackage{braket, mathtools, stmaryrd}

\DeclareMathOperator*{\polylog}{polylog}
\DeclarePairedDelimiter{\abs}{\lvert}{\rvert}
\DeclarePairedDelimiter{\arange}{\llbracket}{\rrbracket}

\usepackage{algorithm}
\usepackage[noend]{algpseudocode}

\usepackage{booktabs}

\begin{document}
\mainmatter              
\title{Improved Methods for $k$-core Community Search}
\titlerunning{Improved $k$-core community search}  
%
\author{Ian Chen\inst{1} \and
    Haotian Yi\inst{1} \and
    Arun Sharma\inst{2} \and
    George Chacko \inst{1} \and
    Tandy Warnow\inst{1}}
\authorrunning{Chen et al.} 
%
\tocauthor{Ian Chen, Haotian Yi, Arun Sharma, Tandy Warnow, and George Chacko}
\institute{University of Illinois Urbana-Champaign, Urbana IL 61801, USA,\\
    \email{\texttt{\char`\{ianchen3,yi54,chackoge,warnow\char`\}@illinois.edu}},\\
    \and {Ladybug Memory, USA
    \email{
    \texttt{arun@ladybugmem.ai}}}}
\maketitle              

\begin{abstract} Community search based on user-specified query nodes is complementary to community finding or graph clustering.
    Prior work in community search is divided into optimizing for external separateness or internal cohesiveness, which does not scale well networks of over a billion edges. We present SteinerKCore, a new scalable $k$-core based community search algorithm for multi-vertex queries. We also present Par-ShellStruct, a parallel algorithm for building the ShellStruct data structure used for $k$-core community search. We show that our implementations in Icebug, an open-source toolkit for large-scale network analysis, are both more efficient and more scalable than comparative tools, being able to perform on a benchmark network of 273M and 5.1B edges using just $64$GB RAM and under $4$ hours runtime with 16 CPUs.
\end{abstract}

\section{Introduction}

Community structure reflects modular organization at the meso-scale of networks. Communities are subsets of the network that are internally cohesive (with relatively high edge density and connectivity) as well as having few edges to the remaining vertices.

A number of algorithmic techniques have been developed to discover community structure.  Of these, community detection and community search are complementary approaches. The former, which can be thought of as a top-down approach, divides the network nodes into disjoint communities. Community search, in contrast, is given a  subset of nodes and finds a single community for that specific set of nodes~\cite{sozio2010communitysearch}.
Community search has multiple advantages over community detection, including reduced computational effort and potentially improved accuracy for finding communities in some cases (e.g., when the true community structure has overlapping communities).
Community search also has natural applications, such as when the interest is in analyzing how the community around a specific node or set of nodes changes over time in a dynamic graph.

The $k$-core community search problem is a well-studied approach in community search.
A $k$-core in a graph $G=(V,E)$ is a maximal subset $V_0 \subset V$ such that the subgraph of $G$ induced by $V_0$ is connected and every node in $V_0$ has at least $k$ neighbors in $V_0$.
Thus, the $1$-cores of a graph are its connected components that have more than a single node.
The $k$-core community search problem is defined as follows: 
Given  a network $G=(V,E)$ and a query set $Q \subset V$, the objective is a $k$-core containing $Q$ for the largest possible $k$.
Since $V_0$ is maximal, the solution is always unique for every query $Q$.
The core number of the query set $Q \subseteq V$, which we also refer to as its coreness, is the largest $k$ such that $Q$ is contained in a $k$-core.

As shown in \cite{fang2019survey}, $k$-core community search can be implemented in near-linear time, making it generally faster and more scalable than many other community search approaches.
In addition, $k$-core communities are more cohesive  than communities found using many alternative approaches, such as optimizing for conductance~\cite{andersen2006local,lin2025effective-etal}.
There are several existing algorithmic approaches to $k$-core community search, including global search, local search, and index-based approaches  well established techniques. 

The \textit{global search} approach~\cite{sozio2010communitysearch} peels the graph in layers, each layer corresponding to vertices with a particular coreness, and returns the  subgraph remaining after the peeling process.
This approach can be implemented efficiently in linear time for a query.
CSK~\cite{fang2019survey,pailodi2024comparison} is one such implementation of global search for the single-vertex case.
As the process of peeling the entire graph to the output community must be repeated for every query, global search is not suited for the regime with many queries.

An alternative approach, referred to as \textit{local search}~\cite{cui2014local}, tries to improve the efficiency of global search by starting the peeling process from a subgraph containing the output.
It searches outwards from the  single-node query until it can prove the subgraph encapsulates the output community.
This procedure can be implemented in linear time for a query.
However, it is limited to only single-vertex queries, and it may be difficult to find subgraphs that are significantly smaller than $G$.

In contrast, \textit{index-based} approaches precompute a data structure that assists in answering queries more efficiently or effectively.
ShellStruct~\cite{barbieri2015efficient} is a tree-based data structure, where nodes correspond to $k$-cores of $G$ and edges are between nodes corresponding to nested subsets; this data structure
can be represented in linear space and built in near-linear time~\cite{fang2016effective}.
Answering community search queries using the ShellStruct tree can then be done in time proportional to the output size, which is clearly optimal. However, building the data structure may be computationally expensive on large graphs.

Although global search, local search, and index-based approaches have good asymptotic complexity, their scalability and computational performance on very large networks have not been sufficiently evaluated. Furthermore, some methods (such as CSK) only work for single-node queries.
\vspace{.05in}

\noindent
\textit{Our contributions:}
We present two new methods (codes available in open-source form at \cite{ian-github}) with excellent computational performance and scalability for $k$-core community search.
The first, SteinerKCore, is designed for the case where there are only a few queries.
The second method is Par-ShellStruct, which provides a parallel algorithm to compute the ShellStruct data structure and then uses it to find $k$-cores for the input queries; this method is designed for the case when the input is a large number of queries.
We also study LocalKCore, an approach we developed based on local search; however, LocalKCore was not as performant as SteinerKCore or Par-ShellStruct, and so is not presented here.



Our study shows that SteinerKCore and Par-ShellStruct are computationally efficient and scalable and can be used for community search on a network as large as $\sim$273 million nodes and $\sim$5 billion edges.
We also compare SteinerKCore and Par-Shellstruct to CSK and ShellStruct, two prior state-of-the-art methods, and establish that SteinerKCore and Par-Shellstruct are faster and can scale to larger networks. 
Due to space limitations, some results are shown in the GitHub site \cite{ian-github}, which also has the open-source codes for our methods.



\section{Our New Methods}

In this section, we  present two methods that return the unique solution to the $k$-core community search problem: SteinerKCore and Par-ShellStruct. Both these methods  use \textit{Icebug} \cite{icebug}, a high-performance graph analysis library forked from NetworKit \cite{staudt2014networkit} that is designed to process massive networks and large-scale graph data while balancing CPU and memory constraints.


For all work in this manuscript, we  assume the input graph $G$ is connected, simple, undirected, and unweighted; the extension to graphs with multiple components is straightforward.

In describing the algorithms and analyzing the runtimes, we use some new notation.
We denote key-value data structures, e.g. dictionaries or priority queues, as $\set{ \text{key} \mapsto \text{value} }$.
The operations GetMaxKey$()$/PopMaxValue$()$ retrieve/extract the maximum key/value in a priority queue.

We let $\arange{n} = \set{1, 2, \ldots, n}$ and
$\alpha(n)$ denote the inverse Ackermann function, which grows extremely slowly and is $\le 5$ in practice (i.e., when $n$ is the number of atoms in the universe). 


\subsection{SteinerKCore}

We present SteinerKCore (named after its relationship to the Steiner max-min  tree problem \cite{chiang1990global}).

\begin{algorithm}[ht]
    \caption{SteinerKCore: given $G = (V, E)$, $Q_1, \ldots, Q_{\ell}$ queries, return $C_i$ community for each $Q_i$, $i \in \arange{\ell}$ solving the $k$-core community search problem}
    \label{alg:steinerkcore}
    \begin{algorithmic}[1]
        \Function{SteinerKCore}{}
            \State Pre-compute $\texttt{coreness}$, the core decomposition of $G$
            \State $UF \gets \text{UnionFind}(V)$, $PQ \gets \set{\texttt{coreness}(q) \mapsto q \mid q \in Q_i, i \in \arange{\ell}}$, $Seen \gets \emptyset$
            \For{$k$ from $\abs{V} - 1$ to $1$}
                \State $R \gets \emptyset$ \Comment{$R$ stores all queries that are completed in this round}
                \While{$PQ.\text{GetMaxKey}() = k$: set $v \gets PQ.\text{PopMaxValue}()$}
                    \State $Seen \gets Seen \cup \set{v}$
                    \For{$u \in N(v)$ s.t. $\texttt{coreness}(u) \ge k$}
                        \If{$UF.\text{Union}(u, v)$ and $Q_i$ becomes connected from edge $uv$}
                            \State add $Q_i$ to $R$ \Comment{mark as completed}
                        \EndIf
                        \If{$u \not \in Seen$}
                            \State $PQ \gets PQ \cup \set{\min(k, \texttt{coreness}(u)) \mapsto u}$
                        \EndIf
                    \EndFor
                \EndWhile
                \State \textbf{output} $C_i \gets UF.\text{ComponentOf}(Q_i)$ for each $Q_i \in R$
            \EndFor
        \EndFunction
    \end{algorithmic}
\end{algorithm}

The Steiner max-min tree (SMMT) problem returns for an instance $G', Q$, where $G'$ is a weighted graph and $Q$ is a subset of nodes, a tree connecting $Q$ maximizing the min-weight of any edge in the tree~\cite{chiang1990global}.
This is related to the community search problem because the set of vertices in the (unique) largest SMMT for $Q$ is exactly the solution to the $k$-core community search problem, if the weight of an edge $uv$ is the minimum core number of the endpoints.

The pseudo-code for SteinerKCore is provided in Alg.~\ref{alg:steinerkcore}.
Note that SteinerKCore has two stages: the first stage is offline, where it  computes the core decomposition and the second stage is online, where it computes the $k$-core community for every query in the input.
SteinerKCore builds a subgraph incrementally, adding vertices in decreasing order of core number; vertices of equal core number form a round. The community for $Q_i$ is the vertex set of its component at the end of the first round in which $Q_i$ is connected. Components are maintained in a union-find augmented with a per-component linked list (for enumerating the component) and a count of query vertices (for detecting when a query becomes connected).

\begin{theorem}
    Given a graph $G = (V, E)$, and $\ell$ queries $Q_1, \ldots, Q_{\ell}$, SteinerKCore (Alg. \ref{alg:steinerkcore}) answers all $\ell$ queries in time $O(\abs{E} \alpha(\abs{V}) + L \log \ell)$, where $L$ is the output size.
\end{theorem}

\subsection{Par-ShellStruct}

We present Par-ShellStruct (Alg. \ref{alg:par-shellstruct}), the first method for building the ShellStruct tree in parallel, which can later be used for answering community search queries in optimal time.
Like SteinerKCore, Par-ShellStruct  has two stages.  The first stage is an offline step that computes the core decomposition and builds the ShellStruct tree and a lowest common ancestor (LCA) table~\cite{bender2000lca}.
The second stage is  online, and is where it finds $k$-core communities for the queries in the input. 
For this second stage, given a single query, it finds all graph vertices in the subtree of the lowest common ancestor (LCA) in the ShellStruct tree for the  vertices in the query.
Thus, when we report the runtime of using Par-ShellStruct given input queries and a network, we mean the total time to construct the ShellStruct tree and then answer the queries.

We parallelize the procedure described in \cite{fang2016effective}, based on the union-find structure over $V$: for $v_1, v_2 \in V$, $\texttt{find}(v_1)$ returns the name of the set containing $v$, and $\texttt{union}(v_1, v_2)$ merges the two sets, returning whether $v_1$ and $v_2$ were already in the same set. 

Par-ShellStruct  builds the ShellStruct data structure from the bottom up by processing each $k$-shell (i.e., all vertices with core number exactly $k$) in bulk, with five substeps (lines 6-10) that are each parallelizable.
In order to avoid data races within each parallel substep, we implement a variant of union find supporting concurrent operations according to suggestions in \cite{alistarh2019search}, namely union-by-rank, plain reads/plain writes, path halving, and immediate parent check.

\begin{algorithm}[ht]
    \caption{Par-ShellStruct: given graph $G = (V, E)$, it builds the ShellStruct index in parallel. $S_k$ denotes the $k$-shell. }
    \label{alg:par-shellstruct}
    \begin{algorithmic}[1]
        \Function{Par-ShellStruct}{}
            \State \Comment{each node in the ShellStruct tree is denoted as $T_{v,k}$, $v \in V$, $k \in \arange{\abs{V}}$; vertices in $G$ are assigned to a unique node of the tree, and nodes may have a parent}
            \State Compute $\texttt{coreness}$, the core decomposition of $G$
            \State Initialize $UF \gets \text{UnionFind}(V)$
            \For{$S_k$ the non-empty $k$-shell in descending $k$}
                \State \textbf{parinit} $P_k \gets \set{ T_{UF.\text{Find}(u), \texttt{coreness}(u)} \mid v \in S_k, uv \in E, \texttt{coreness}(u) > k }$
                \State \textbf{parfor} $uv \in \set{ uv \in E \mid u \in S_k, \texttt{coreness}(v) \ge k }$ \textbf{do} $UF.\text{Union}(u, v)$
                \State \textbf{parfor} $v \in \set{ UF.\text{Find}(u) \mid u \in S_k }$ \textbf{do} create new node $T_{v, k}$
                \State \textbf{parfor} $u \in S_k $ \textbf{do} assign $u$ to $T_{UF.\text{Find}(u), k}$
                \State \textbf{parfor} $T_{r, k'} \in P_k$ $\textbf{do}$ set parent of $T_{r, k'}$ to $T_{UF.\text{Find}(r), k}$
            \EndFor
            \State $k_{\text{min}} \gets \min \set{ k \mid S_k \ne \emptyset}$; \Return unique ShellStruct node with value $k_{\text{min}}$
        \EndFunction
    \end{algorithmic}
\end{algorithm}

\begin{theorem}
    Par-ShellStruct (Alg. \ref{alg:par-shellstruct}) is work-efficient, i.e. matches the work of the best sequential algorithm.
    It can be implemented with $O(\rho(G) \polylog(\abs{V}))$ span with high probability in the Parallel RAM model, where $\rho(G) < \abs{V}$ is the peeling-complexity defined in \cite{dhulipala2020graph}.
\end{theorem}



\section{Experimental Study}

\subsection{Study Design}

\noindent
\textit{Methods.}
We study our new methods, SteinerKCore and Par-ShellStruct, as well as two prior state-of-the-art methods, CSK (an efficient implementation of the global search strategy from \cite{fang2016effective}) and a protocol based on building the ShellStruct \cite{barbieri2015efficient} datastructure and then using it to answer queries (just as we do for Par-ShellStruct); see the Introduction section for additional details about these prior methods.



\begin{table}[h!]
    \caption{
        \textbf{Networks used in the experimental study.} For each network, we provide  a reference to an online source and empirical statistics (the number of nodes, the number of edges, and the average degree, each computed after removing parallel edges and self-loops).
    }
    \centering
    \begin{tabular}{lrrr}
        \hline
        Network                                              & \# Nodes    & \# Edges      & avg deg \\
        \hline
        \textbf{Training networks}                                                                   \\
        \texttt{abm14}~\cite{park2026modeling}               & 13,926,21   & 581,472,875   & 83.51   \\
        \texttt{CEN}~\cite{park2024data}                     & 13,989,43   & 92,051,051    & 13.16   \\

        \hline
        \textbf{Testing networks}                                                                    \\
        \texttt{LiveJournal}~\cite{peixoto2020netzschleuder} & 4,846,609   & 42,851,237    & 17.68   \\
        \texttt{Bitcoin}~\cite{peixoto2020netzschleuder}     & 6,297,539   & 15,464,723    & 4.91    \\
        \texttt{Wikipedia}~\cite{peixoto2020netzschleuder}   & 13,593,03   & 334,591,525   & 49.23   \\

        \texttt{MS-Concept}~\cite{peixoto2020netzschleuder}  & 16,936,66   & 33,354,319    & 3.94    \\
        \texttt{DBpedia}~\cite{peixoto2020netzschleuder}     & 18,268,99   & 126,890,209   & 13.89   \\
        \texttt{Twitter}~\cite{peixoto2020netzschleuder}     & 41,652,23   & 1,202,513,046 & 57.74   \\
        \texttt{Friendster}~\cite{peixoto2020netzschleuder}  & 65,608,36   & 1,806,067,135 & 55.05   \\
        \texttt{abm272mf}~\cite{abmnetworks2026}            & 272,739,486 & 5,170,223,924 & 37.90   \\
        \hline
    \end{tabular}

    \label{tab:networks}
\end{table}
\vspace{.05in}

\noindent
\textit{Networks.} 
We use a collection of ten networks (see Table \ref{tab:networks}).
We developed three of these networks: abm14, abm272mf, and the CEN.
The abm14 and abm272mf networks were created by agent-based models (ABMs) designed to simulate the growth of a citation network over a period of multiple decades \cite{park2026modeling,abmnetworks2026}.
The CEN (Curated Exosome Network) is a real-world citation network derived from the exosome biology literature \cite{park2024data}. The remaining networks are all real-world networks   from the Netzschleuder repository~\cite{peixoto2020netzschleuder} that have at least $4$ million nodes). All networks were cleaned to remove parallel edges and self-loops and then re-indexed to enforce continuous node IDs. After cleaning, the networks range  from
$\sim$4.8 million to $\sim$273 million nodes and have $\sim$15 million to $\sim$ 5 billion edges.
The  largest network is abm272mf~\cite{abmnetworks2026} which has 272,739,486  (i.e., $\sim$273M) nodes and   $\sim$5.1 billion edges.


\vspace{.05in}

\noindent
\textit{Computing resources.}
Unless otherwise specified, we give each method a maximum of 4 hours of runtime, 16 CPU cores, and 128 GB RAM.
All experiments are run on the Illinois Campus Cluster, a cluster of heterogeneous compute nodes.

\vspace{.05in}

\noindent
\textit{Query generation}
We expect typical workflows to query nodes that are influential in the network.
Thus, we generated queries by sampling nodes that are in the same connected component of $G$ (without replacement) within the top 1\% by degree. 

\vspace{0.05in}
\noindent
\textit{Evaluation criteria}
Since all tested methods solve the $k$-core community search problem exactly, we only evaluate with respect to computational performance (runtime and peak memory usage) and scalability.

\vspace{.1in}

\noindent
\textit{Experiments.}

\begin{itemize}
    \item Experiment 1: Algorithmic design.
          1a: We determine the fastest code for computing the core decomposition.
          1b: We determine which of our methods is best for a single query. These experiments are performed on the two training networks.
    \item Experiment 2: Comparing our $k$-core community search methods to prior methods.
          We compare our proposed community search methods with the best existing methods on the testing networks.
    \item Experiment 3: Evaluating  computational performance and scalability.
          3a: We evaluate how our methods scale when increasing CPU count.
          3b: We evaluate how our methods scale when increasing network size. These experiments are performed on the testing networks.
\end{itemize}

\subsection{Results}

\vspace{.05in}

\noindent
\textit{Experiment 1:}
SteinerKCore and Par-Shellstruct have as their first step computing the core decomposition. Therefore, in designing these methods, we first evaluate the options for this step, using six different implementations: Icebug~\cite{icebug}, UCR~\cite{liu2025parallel}, GBBS~\cite{dhulipala2020graph}, NK~\cite{staudt2014networkit}, PKC~\cite{kabir2017parallel}, and Ladybug~\cite{ladybug}.  We use the two training networks for this experiment.

\begin{figure}[ht]
    \centering
    \includegraphics[width=0.8\textwidth]{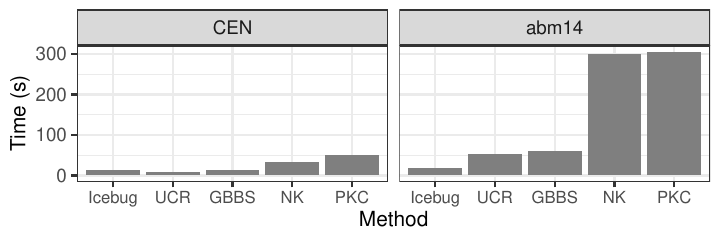}
    \caption{
        \textbf{Experiment 1: core decomposition.}
        We show the runtime of five external core decomposition implementations on two networks (\texttt{CEN}, \texttt{abm14}) with 14M nodes; \texttt{ABM} has $6\times$ more edges than \texttt{CEN}.
        Ladybug times out on both networks and is not shown.
        Based on these results, we conclude Icebug to be the preferred implementation --- it is second best on \texttt{CEN} but best on \texttt{abm14}}.
    \label{fig:train-core-decomp}
\end{figure}

On these two networks (each of which has $\sim$14 million nodes), Icebug and UCR are  the  fastest methods,  with GBBS a close third (Figure \ref{fig:train-core-decomp}).
Icebug and UCR are very close on the CEN but Icebug has a substantial advantage over UCR on abm14.
Therefore, we select Icebug as the code for computing the core decomposition in our study.


\noindent
\textit{Experiment 2:}
We compare  SteinerKCore and Par-ShellStruct to prior SOTA methods, Shellstruct and CSK.
We use Icebug's core decomposition routine as the first step for each method.
We compare the methods on the testing networks, varying the query size (from 1 to 20 nodes) and number of queries (from  1 to 100);  query sizes of 5 and 10 are similar and shown in the  supplementary materials on the GitHub site.

\begin{figure}[p]
    \centering
    \includegraphics{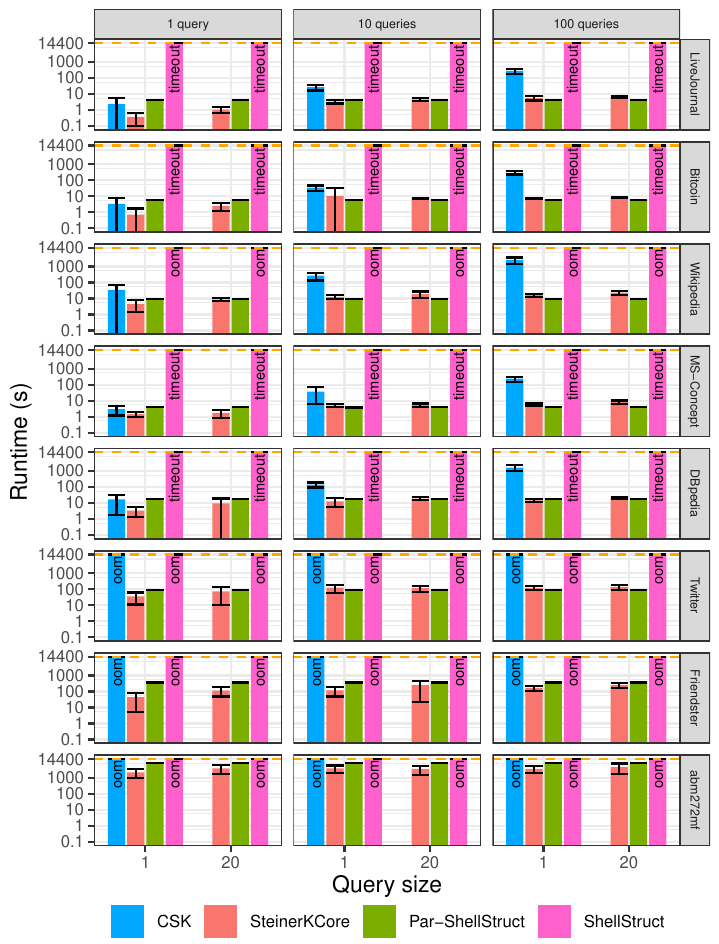}
    \caption{
        \textbf{Experiment 2: Runtimes  of four community search methods (CSK, SteinerKCore, Par-ShellStruct, and ShellStruct) on testing networks given 16 cores.}
        Runtimes do not include  the time for computing the core decomposition, as all use the same technique.
         Each row indicates a network, sorted by  number of vertices (from 5M to 273M).
        Each bar shows average (error bars are std dev) for 20 batches of queries, each with 1--100  queries with 1--20 vertices per query.
        CSK only takes as input single-vertex queries and so is not shown for queries with 20 vertices.
        Failures are out-of-memory (oom) or timeout.
    }
    \label{fig:test-commsearch}
\end{figure}

The results for this experiment (Figure \ref{fig:test-commsearch})  do not include CSK for the multi-node  query case, as it is not designed for such inputs. 
For queries with single nodes,
CSK runs out of memory on the three largest networks.
Even given a single query,
ShellStruct does not complete on any of these networks, either running out of memory or exceeding the 4-hour time limit.
Furthermore, on the inputs on which the methods complete, CSK and ShellStruct are slower than both SteinerKCore  and Par-ShellStruct.
Thus, SteinerKCore and Par-ShellStruct have superior performance on these inputs.

SteinerKCore is consistently faster than Par-ShellStruct for the single query case (whether it contains  1 or 20 vertices), while Par-ShellStruct is generally but not always faster than SteinerKCore when there are 100 queries.
This generally confirms our expectation that Par-ShellStruct would be better suited for the many-query case and  SteinerKCore better suited for the single query case.
On the other hand, given  10 queries, the relative performance is mixed, sometimes favoring  SteinerKCore and sometimes favoring Par-ShellStruct.

\vspace{.1in}
\noindent
\textit{Experiment 3:}
In Experiment 3a, 
we evaluate how SteinerKCore and Par-ShellStruct scale with the number of CPU cores.
We use a dedicated cluster, Folkvangr, which has 256 GB for this experiment. 
We examined runtimes for each of the testing networks; results on three representative networks (a ``small" network, \texttt{LiveJournal}, with 4.8M nodes and 15M edges; a  ``medium" network, \texttt{Wikipedia}, with 13.6M nodes and 334M edges; and  a ``large" network, \texttt{Friendster}, with  65.6M nodes and 1.8B edges)  are shown in Figure \ref{fig:strongscaling}, with the remaining networks shown in the supplementary materials.

\begin{figure}[ht]
    \centering
    \includegraphics[width=0.8\textwidth]{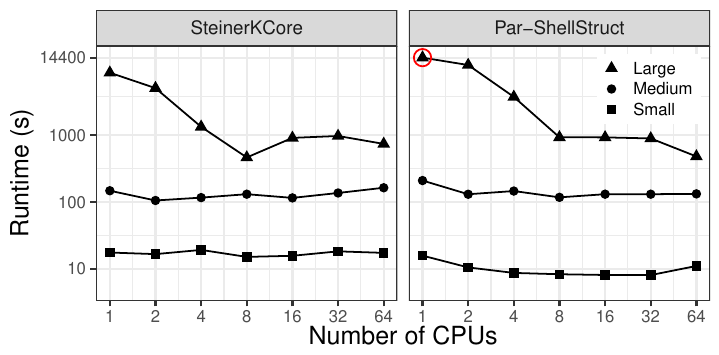}
    \caption{
        \textbf{Experiment 3a: strong-scaling.}
        Each facet shows the runtime of SteinerKCore and Par-ShellStruct, our community search methods.
        For 1, 2, 4, 8, 16, 32, and 64 CPUs, we plot the mean runtime among 5 batches of 32 5-vertex queries.
        We show a small (\texttt{LiveJournal}, 4.8 million nodes and 15M edges), medium (\texttt{Wikipedia}, 13.6 million nodes and 334M edges), and large (\texttt{Friendster}, 65.6 million nodes and 1.8B edges) network.
        Par-ShellStruct times out on the large network at 1 CPU.
        Results on the remaining networks are shown in the supplementary materials on the GitHub site.
    }
    \label{fig:strongscaling}
\end{figure}

The online phase of both methods (i.e. the query processing step) have limited parallelism; we only parallelize across queries.
We select 32 5-vertex query sets and record the runtime; we repeat this five times and average the runtimes across the batches.
Increasing the number of cores tends to reduce the runtime for both methods on the large network  but less so on the medium    and small networks (Figure \ref{fig:strongscaling}).   
SteinerKCore shows less benefit from scaling than  Par-ShellStruct.

\begin{figure}[ht]
    \centering
    \includegraphics[width=0.8\textwidth]{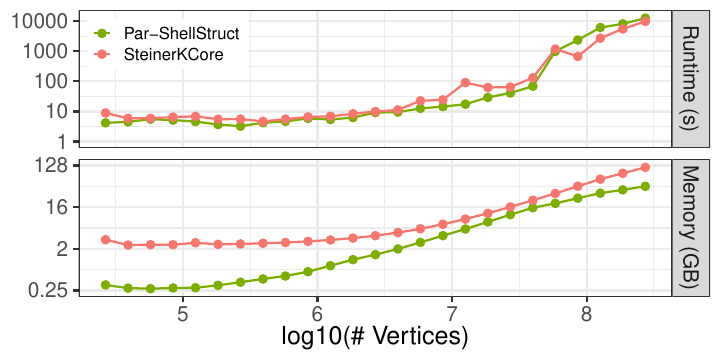}
    \caption{
        \textbf{Experiment 3b: scaling with input size.}
        We plot the total runtime and peak memory usage of community search methods given 5 batches of 32 5-vertex queries on increasing network sizes using time-snaps of the abm272mf network ($\sim$273 million nodes and $\sim$5.1 billion edges).
    }
    \label{fig:scaling-abm272}
\end{figure}

In Experiment 3b, we evaluate the time and memory usage as we scale the input size, keeping 16 CPU cores.
For this experiment, we explore subnetworks of our largest network, abm272mf.
This network was generated by an agent-based model implemented in the SASCA-ReSA \cite{sascaresa_fi_fork} generator, which simulates the growth of a citation network from an initial seed network, adding new nodes (publications) with citations (edges) to prior publications each year, over the 20-year period from 2026 (26k nodes) to 2046 (273M nodes).

Every five years, we generate 5 batches of 32 5-vertex community search queries, and report the mean time and memory to answer each batch (Fig. \ref{fig:scaling-abm272}).
On the final network, SteinerKCore finishes in 2 hours 45 mins using 116GB memory, while Par-ShellStruct takes longer, finishing in 3 hours 31 minutes but only using 45GB memory.
We note that the memory cost of SteinerKCore is due to $\min(\text{number of CPUs}, 32)$ simultaneous processes being spawned to parallelize across queries --- the peak memory usage for any individual process is 39GB; this indicates a trade-off between memory usage and runtime.
We also note that approximately half the time for each method is spent on computing the core decomposition.

Both time and memory follow roughly piecewise-linear trends on a log-log plot.
Above $10^6$ nodes, memory scales as $\abs{V}^{0.77}$ for Par-ShellStruct and $\abs{V}^{0.71}$ for SteinerKCore.
Runtime is sublinear below $5 \times 10^7$ nodes, at $\abs{V}^{0.33}$ and $\abs{V}^{0.39}$ respectively, but superlinear beyond it, at $\abs{V}^{1.64}$ and $\abs{V}^{1.65}$.

\subsection{Comparing SteinerKCore and Par-ShellStruct}
The results shown establish that  SteinerKCore and Par-ShellStruct are faster and more scalable than the previous methods we examined, ShellStruct and CSK.
Here we compare SteinerKCore and Par-Shellstruct.

Experiment 2 shows that SteinerKCore generally has an advantage over Par-ShellStruct for runtime when the number of queries is small enough (i.e., consistently faster for 1 query) but Par-ShellStruct is faster at 100  queries, with mixed results on  10 queries.
Experiment 3 shows that the two methods have similar runtimes on the subgraphs of the largest network we studied (5 queries), but Par-ShellStruct uses much less memory (at  most 45GB whereas SteinerKCore used 116GB).
Both exhibit strong scaling on the large network we explored (\texttt{Friendster}) but not on the other two networks.
On the large network, Par-ShellStruct is slower than SteinerKCore whenever the number of cores is at most 8, and is only faster at 64 cores.  

Hence, the choice between the two methods depends on many factors, including the number of queries (with a very small number favoring SteinerKCore  and a large number favoring Par-ShellStruct) and whether  available memory is limited (favoring Par-ShellStruct).

\section{Conclusion}

Community search has the potential for greater personalization, being more efficient, and able to scale to larger networks than community detection.
In this study, we present two new implementations for $k$-core community search, SteinerKCore and Par-ShellStruct, and show that they have better computational performance on large networks (up to 273 million nodes) than state-of-the-art methods. We also show that these methods can complete quickly with limited resources, while competing methods fail due to memory requirements. Our new methods represent an advance in practical  tools for community search on large networks.

\section*{Use of Artificial Intelligence}

The authors declare the interactive use of assistive AI~\cite{anthropic2026} for generation of software codes and experimental scripts. The text of the manuscript was written, and approved by the authors who acknowledge full responsibility for the content and conclusions presented.

%
%

\bibliographystyle{splncs04}
\bibliography{refs}
\end{document}